\documentclass[journal,12pt,onecolumn]{IEEEtran}
\usepackage{multicol,lipsum,stfloats}
\usepackage{setspace}
\usepackage{color}
\usepackage{soul}
\def\x{{\mathbf x}}
\def\z{{\mathbf z}}
\def\y{{\mathbf y}}
\def\H{{\mathbf H}}
\def\R{{\mathbf R}}
\def\A{{\mathbf A}}
\def\U{{\mathbf U}}
\def\Q{{\mathbf Q}}
\def\h{{\mathbf h}}
\def\b{{\beta}}
\def\g{{\gamma}}
\def\d{{\delta}}
\def\e{{\eta}}

\ifCLASSINFOpdf
\else
\fi
\usepackage{epsfig}
\usepackage{epstopdf}

\usepackage{pbox}

\usepackage[cmex10]{amsmath}
\usepackage{amsfonts}
\usepackage{subfigure}

\usepackage{caption}
\begin{document}
%
\title{Sparsity-aware sphere decoding: Algorithms and complexity analysis}
%
%
%
\author{Somsubhra Barik and Haris Vikalo,~\IEEEmembership{Member,~IEEE}
 \thanks{The authors are with the Department
 of Electrical and Computer Engineering, The University of Texas at Austin, Austin,
 TX, 78712 USA (e-mail: sbarik@utexas.edu, hvikalo@ece.utexas.edu).
The preliminary work was presented at the 2013 IEEE International Conference
on Acoustic, Speech and Signal Processing \cite{icassp2013}.}
}

\maketitle

\begin{abstract}
Integer least-squares problems, concerned with solving a system of equations where the
components of the unknown vector are integer-valued, arise in a wide range of applications. 
In many scenarios the unknown vector is sparse, i.e., a large fraction of its entries are zero.
Examples include applications in wireless communications, digital fingerprinting, and
array-comparative genomic hybridization systems. Sphere decoding, commonly used for
solving integer least-squares problems, can utilize the knowledge about sparsity of the
unknown vector to perform computationally efficient search for the solution. In this paper, 
we formulate and analyze the sparsity-aware sphere decoding algorithm that imposes 
$\ell_0$-norm constraint on the admissible solution. Analytical expressions for the expected 
complexity of the algorithm for alphabets typical of sparse channel estimation and source 
allocation applications are derived and validated through extensive simulations. The results 
demonstrate superior performance and speed of sparsity-aware sphere decoder compared 
to the conventional sparsity-unaware sphere decoding algorithm. Moreover, variance of
the complexity of the sparsity-aware sphere decoding algorithm for binary alphabets is
derived. The search space of the proposed algorithm can be further reduced by imposing
lower bounds on the value of the objective function. The algorithm is modified to 
allow for such a lower bounding technique and simulations illustrating efficacy of the method 
are presented. Performance of the algorithm is demonstrated in an application to 
sparse channel estimation, where it is shown that sparsity-aware sphere decoder
performs close to theoretical lower limits. 

\end{abstract}


 \ifCLASSOPTIONpeerreview
 \begin{center} \bfseries EDICS Category: SPC-DETC \end{center}
 \fi
%
\IEEEpeerreviewmaketitle

\section{Introduction}
\begin{figure}[!t]
\centering
\includegraphics[width=4in]{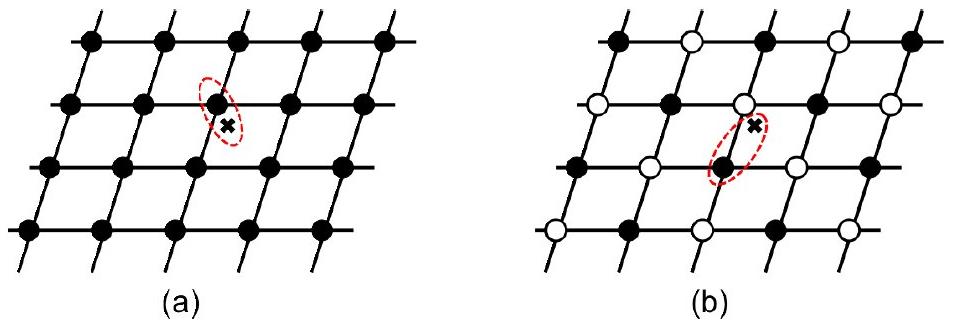}
\caption{\small Illustration of the search for the closest lattice point. The black dots in both 
subfigures denote lattice points $\H\x$. ${\bf \times}$ marks the 
observed vector $\y$. Subfigure (a) illustrates the scenario where all lattice points are
admissible while (b) illustrates the scenario where the unknown vector is sparse and thus
not all lattice point can be the solution.}
\label{lattice_fig}
\end{figure}

Given a matrix ${\bf H} \in  \mathbb{R}^{n \times m}$ and a vector $\y \in  \mathbb{R}^{n \times 1}$, 
the integer least-squares (ILS) problem is concerned with finding an $m$-dimensional vector 
${\bf x}^{\star}$ comprising integer entries such that 
\begin{eqnarray}
& {\bf x}^{\star} ~= ~\arg\underset{{\bf x} \in \mathbb{Z}^m}{\min}~~\|\y  -\H\x\|_2^2  \label{ils}, & 
\end{eqnarray} 
where $\mathbb{Z}^m$ denotes the $m$-dimensional integer lattice. In many applications, 
the unknown 
vector $\x$ belongs to a finite $m$-dimensional subset $\mathcal{D}_L^m$ of the infinite lattice 
$\mathbb{Z}^m$ such that $\mathcal{D}_L^m$ has $L$ elements per dimension, i.e., each component 
of the unknown vector $\x$ can take one of $L$ possible discrete values from the set 
$\mathcal{D}_L \subset \mathbb{Z}$. In multi-antenna communication systems, for
instance, $\x \in \mathcal{D}_L^m$ is the transmitted symbol while the received signal
$\y = {\bf H} \x + \boldsymbol\nu$ is perturbed by the additive noise $\boldsymbol\nu$
(hence, $\boldsymbol\nu$ renders the system of equations $\y \approx {\bf H} \x$ 
inconsistent). Note that the symbols in $\mathcal{D}_L^m$ form an $m$-dimensional rectangular 
lattice and, therefore, ${\bf Hx}$ belongs to an $n$-dimensional lattice skewed in the direction of the 
eigenvectors of ${\bf H}$. An integer least-squares problem can be interpreted as the search for the nearest
 point in a given lattice, commonly referred to as the {\em closest lattice point} problem \cite{agrell}. 
Geometric interpretation of the integer least-squares problem as the closest lattice point search is 
illustrated in Fig. \ref{lattice_fig}(a). 

Integer least-squares problems with sparse solutions arise in a broad range of applications including multi-user detection
in wireless communication systems \cite{MUD}, sparse array processing \cite{array}, collusion-resistant digital fingerprinting 
\cite{collu1}, and array-comparative genomic hybridization
microarrays \cite{microarrays}.
Formally, a sparse integer least-squares problem can be stated as the cardinality constrained 
optimization
\vspace{-1.2mm}
\begin{eqnarray}
\min_{\x \in \mathcal{D}_L^m} ~~~~~~\|\y  -\H\x\|_2^2  \label{firsteq} && \\
\mbox{subject to  } ~~~\|\x\|_0 ~\leq~ l, && \nonumber
\end{eqnarray}
where $\|\cdot\|_0$ denotes the $\ell_0$ norm of its argument and $l$ is an upper 
bound on the number of non-zero entries of $\x$. 
In this paper, we restrict our attention to the case where $n \geq m$
(all of the aforementioned applications fall in this category). 
Note that (\ref{firsteq}) can be interpreted as a search for the lattice point $\H\x$ 
closest to the given point ${\bf y}$ in a sparse lattice, where the sparsity constraint is
explicitly stated as $\{\x \in \mathcal{D}_L^m  ~|~ \|\x\|_0 \leq l\}$. 
Search over a sparse lattice is illustrated in Fig. \ref{lattice_fig}(b).

Finding the exact solution to (\ref{firsteq}) is computationally intensive (in fact,
the closest lattice point problem is known to be NP hard \cite{NP}). 
The sphere decoding algorithm, developed by Fincke and Pohst \cite{fincke}, 
efficiently solves the ILS problem and provides optimal solution 
in a broad range of scenarios \cite{vikalo0,vikalo1,vikalo2}. 
In particular, it was shown in \cite{vikalo1,vikalo2} that if the sphere radius is chosen 
using the statistics of $\boldsymbol\nu = \y-\H\x$, then the 
sphere decoding algorithm has an expected complexity that is practically feasible  
over a wide range of problem parameters.

Recently, several variants of sphere decoder that exploit information about sparsity 
of the unknown vector were proposed \cite{array}, \cite{finite}, \cite{MUD}. In \cite{array},
a modified sphere decoding algorithm with the $\ell_0$-norm constraint relaxed to an
$\ell_1$-norm regularizer was proposed. This scheme is only applicable to non-negative 
alphabets, in which case $\|\x\|_1$ can be decomposed into the sum of the components 
of $\x$. In \cite{finite}, a generalized sphere decoding approach imposing an $\ell_1$ 
constraint on the solution was adopted for sparse integer least-square problems over 
binary alphabets. That work examines the case where $n < m$ and essentially considers
compressed sensing scenario. An $\ell_0$-norm regularized sphere decoder has been  
proposed and studied in \cite{MUD}, where the regularizing parameter $\lambda$ was
chosen to be a function of the prior probabilities of the activity of independent users in 
a multiuser detection environment. In contrast, sphere decoder in the present 
manuscript directly imposes the $\ell_0$-norm constraint on the unknown vector, 
i.e., we perform no relaxation or regularization of the distance metric. Note that the 
closest point search in a sparse lattice has 
previously been studied in \cite{vikalosparse} but sparsity there stems from the fact 
that not all lattice points are valid codewords of linear block codes. 
In \cite{younis}, a sparse integer least-squares problem arising in the context of 
spatial modulation was studied for the special case of symbol vectors with single 
non-zero entry; the method there does not rely on the branch-and-bound search
typically used by sphere decoding.
Note that none of the previous works on characterizing complexity of sphere decoder 
(see, e.g., \cite{jalden}, \cite{seethaler}, \cite{SDinfty} and the references therein)
considered sparse integer least-squares problems except for \cite{younis2}, where
the worst-case complexity of sphere decoder in spatial modulation systems of 
\cite{younis} was studied. Finally, we should also point out 
a considerable amount of related research on compressed sensing, where one is
interested in the recovery of an inherently sparse signal by using potentially far fewer 
measurements than what is typically needed for a signal which is not sparse 
\cite{chen95}-\cite{cand06a}. 

The paper is organized as follows. In Section~II, we review the sphere decoding
algorithm and formalize its sparsity-aware variant. Following the analysis
framework in \cite{vikalo1,vikalo2}, in Section~III we derive the expected
complexity of the sparsity-aware sphere decoder for binary and ternary alphabets
commonly encountered in various applications. The derived analytical expressions
are validated via simulations. Section~IV presents an expression for the variance 
of the complexity of the proposed algorithm for binary alphabets. In Section~V, the algorithm is modified by introducing an 
additional mechanism for pruning nodes in the search tree, leading to a significant 
reduction of the number of points traversed and the overall computational complexity. 
In Section~VI, performance of the algorithm is demonstrated in an application to sparse 
channel estimation.
The paper is summarized in Section~VII.

\section{Algorithms}

In this section, we first briefly review the classical sphere decoding algorithm and then 
discuss its modification that accounts for sparsity of the unknown vector.

\subsection{Sphere decoding algorithm}
\label{sd_overview}

To find the closest point in an $m$-dimensional lattice, sphere decoder performs the search within an 
$m$-dimensional sphere of radius $d$ centered at the given point $\y \in \mathbb{R}^{n}$. In
particular, to solve (\ref{ils}), the sphere decoding algorithm searches over $\x$ such
that $d^2 \geq \|{\bf y} - {\bf Hx}\|^2$. The search procedure is reminiscent of the branch-and-bound 
optimization \cite{murugan}. In a pre-processing step, one typically starts with the $QR$-decomposition 
of $\H$,
\begin{eqnarray}
\H & = & [\Q_1 ~~ \Q_2]\left[ \begin{array}{c}
\R \\
{\bf 0}_{(n-m \times m)}
\end{array}\right],
\end{eqnarray}
where $\R$ is the upper-triangular matrix and $\Q_1$ and $\Q_2$ are composed of the first $m$ 
and the last $n - m$ orthonormal columns of $\Q$, respectively. Therefore, we can rewrite the sphere 
constraint $d^2 \geq \|{\bf y} - {\bf Hx}\|^2 $ as 
\begin{eqnarray}
 d^2 - \|\Q_2^*{\bf y}\|^2 & \geq & \|{\bf z} - \R{\bf x}\|_2^2, \label{sphrcons}
\end{eqnarray}
where ${\bf z} = \Q_1^*{\bf y}$. Now, (\ref{sphrcons}) can be expanded to 
\begin{align}
& d^2 - \|\Q_2^*{\bf y}\|^2  \geq  (z_m - R_{m,m}x_m)^2 
 + (z_{m-1} - R_{m-1,m}x_m - R_{m-1,m-1}x_{m-1})^2 + \cdots & 
\label{SDeq1}
\end{align}
where $x_i$ and $ z_i$ are the $i^{th}$ components of $\x$ and $\z$, respectively, and $R_{i,j}$ 
is the $(i,j)^{th}$ entry of $\R$. Note that the first term on the right-hand side (RHS) of 
(\ref{SDeq1}) depends only on $x_m$, the second term on $\{x_{m}, x_{m-1}\}$, and so on. 
A necessary (but not sufficient) condition for ${\bf Hx}$ to lie inside the sphere is  
$d^2 - \|\Q_2^*{\bf y}\|^2  \geq (z_m - R_{m,m}x_m)^2$. For every $x_m$ satisfying this 
condition, a stronger necessary condition can be found by considering the first two terms on 
the RHS of (\ref{SDeq1}) and imposing a condition that $x_{m-1}$ should satisfy 
to lie within the sphere. One can proceed in a similar fashion to determine conditions for 
$x_{m-2}, \ldots, x_1$, thereby determining all the lattice points that satisfy 
$d^2 \geq \|{\bf y} - {\bf Hx}\|^2_2$. If no point is found within the sphere, its radius is 
increased and the search is repeated. If multiple points satisfying the constraint are found,
then the one yielding the smallest value of the objective function is declared the solution.
Clearly, the choice of the radius is of critical importance for facilitating a computationally 
efficient search. If $d$ is too large, the search complexity may become infeasible, while if 
$d$ is too small, no lattice point will be found within the sphere. To this end, $d$
can be chosen according to the statistics of $\boldsymbol\nu = \y - \H \x$, hence
providing probabilistic guarantee that a point is found inside the sphere \cite{vikalo1}.

\begin{figure*}
\centering
\includegraphics[width=5.5in]{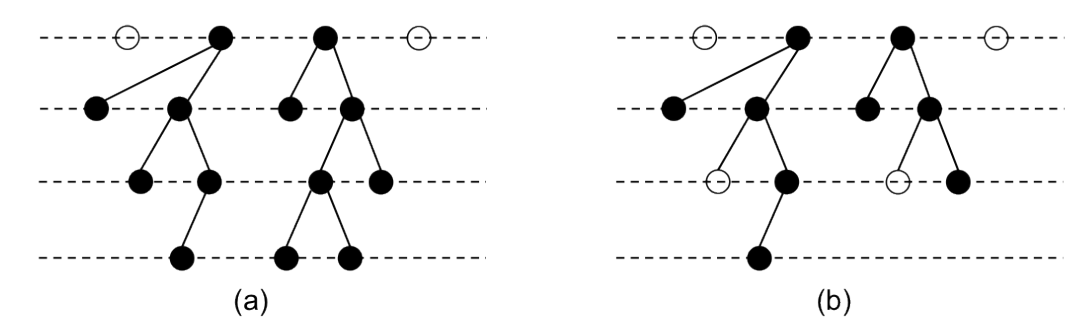}
\caption{\small An illustration of the depth-first tree search of the sphere decoding algorithm 
when (a) the unknown vector is non-sparse and (b) the unknown vector is sparse.}
\label{sphere_tree}
\end{figure*}
The sphere decoding algorithm can be interpreted as a search on a tree, as illustrated in 
Fig. \ref{sphere_tree}(a). The nodes at the $k^{th}$ tree level represent $k$-dimensional 
points $[x_{m-k+1} ~ x_{m-k+2}~ \ldots x_m]^T$ . The algorithm uses aforementioned 
constraints to prune nodes from the tree, keeping only those that belong to the 
$k$-dimensional sphere of radius $d$.  Many variants of the
basic sphere decoding have been developed \cite{Steingrimsson}, \cite{Zhan} 
- 
\cite{Barbero}.

\subsection{Sparsity-aware sphere decoding}
\label{sparseSD}

In scenarios where the unknown vector $\x$ is known to be sparse, imposing $\ell_0$-norm 
constraint on ${\bf x}$ improves the speed of the search since the necessary conditions that 
the components of ${\bf x}$ must satisfy become more restrictive. Clearly, not all lattice 
points within the search sphere satisfy the sparsity constraint and thus fewer points 
need to be examined in each step of the sparsity-aware sphere decoding algorithm.
We impose sparsity constraint on the components of ${\bf x}$ at each node of the search
tree traversed by the algorithm. Note that the number of 
non-zero symbols along the path from the root node to a given node is a measure of the 
sparseness of the $k$-dimensional point associated with that node. Suppose that a node 
at the $k^{th}$ level of the tree satisfies the sphere constraint. Sparsity constraint implies
that, in addition to the node being inside the sphere, the number of non-zero symbols along 
the path leading to the node must not exceed an upper bound $l$. Hence, knowledge
of sparsity allows us to impose more strict conditions on the potential solutions to the
integer least-squares problems, and the number of nodes that the algorithm prunes is
greater than that in the absence of sparsity (or in the absence of the knowledge about sparsity).

\begin{large}
 \vspace{10mm}
\begin{table}\centering
\caption{{\small Sparsity-aware Sphere Decoding Algorithm}}
\vspace{1.5mm}
\begin{tabular}{l}
\hline
 Input: $Q = [Q_1 ~~ Q_2], ~ R, y, ~z = Q_1^*y,$ 
 sphere radius $d$, \\ sparsity constraint $l$. \\
\hline \\
 1. \underline{ Initialize}  $k \leftarrow m$, $d_m^2 \leftarrow d^2- \|Q_2^*y\|^2$, \\
$z_{m|m+1} \leftarrow z_m$,  $l_m \leftarrow 0$ . \\
2. \underline{ Update Interval} ~~
 $UB(x_k) \leftarrow \left \lfloor (d_k + z_{k|k+1})/R_{k,k} \right \rfloor$,  
\\ $LB(x_k) \leftarrow \left \lceil  (-d_k + z_{k|k+1})/R_{k,k} \right \rceil  $,
 $x_k \leftarrow LB(x_k)-1$. \\
3. \underline{ Update $x_k$}   ~~ $x_k \leftarrow x_k+1$.
If $x_k \leq UB(x_k),$ \\ go to  4; else go to 5. \\ 
4. \underline{Check Sparsity} ~~~ If $l_k+ \mathcal{I}_{\{x_k \neq 0\}} \leq l,~$ \\
$l_k \leftarrow l_k + \mathcal{I}_{\{x_k \neq 0\}}$, and go to  6; else go to 3.\\
5. \underline{Increase $k$}   ~~ $k \leftarrow k+1$. 
If  $k=m+1,$ stop; \\
else, $l_k \leftarrow l_k - \mathcal{I}_{\{x_k \neq 0\}}$ and go to 3. \\
6. \underline{Decrease $k$} If $k=1,$ go to 7; else $k \leftarrow k-1$, \\
$z_{k|k+1} \leftarrow z_k - \sum_{j=k+1}^{m}R_{k,j}x_j, $~ \\
$d_k^{2} \leftarrow d_{k+1}^{2} $
$-(z_{k+1|k+2} - R_{k+1,k+1}x_{k+1})^2,$  and go to 2. \\
7. \underline{Solution found}~~ Save $x$ and its distance from $y$, \\
$d_m^{2} - d_1^{2} + (z_1-R_{1,1}~x_1)^2,~ l_k \leftarrow l_k - \mathcal{I}_{\{x_k \neq 0\}}$ \\
and go to 3. 
\\
\hline
\end{tabular}
\label{algo1}
\end{table}
\end{large}



Table \ref{algo1} formalizes the sparsity-aware sphere decoding algorithm with
$\ell_0$-norm constraint imposed on the solution. Note that, in the pseudo-code, 
variable $l_k$ denotes the number of non-zero symbols selected in the first $k-1$ 
levels. This variable is used to impose the sparsity constraint in Step 4 of the 
algorithm (calculating $l_k$ and imposing the sparsity constraint requires only a 
minor increase in the number of computations per node). Whenever the algorithm 
backtracks up a level in the tree, the value of $l_k$ is adjusted to reflect sparseness of the current 
node (as indicated in Steps 5 and 7 of the algorithm). The depth-first tree search 
of the sparsity-aware sphere decoding algorithm is illustrated in Figure \ref{sphere_tree}(b),
where fewer points survive pruning than in Figure \ref{sphere_tree}(a).

{\em Remark 1:} The algorithm in Table \ref{algo1} relies on the original Fincke-Pohst
strategy for conducting the tree search \cite{fincke}. There exist more efficient implementations
such as the so-called Schnorr-Euchner variant of sphere decoding where the search in each 
dimension $k$ starts from the middle of the feasible interval for $x_k$ and proceeds to explore
remaining points in the interval in a ``zig-zag" fashion (for details see, e.g., \cite{agrell}).
For computational benefits, one should combine such a search strategy with radius update, i.e., 
as soon as a valid point $\x_{in}$ inside the sphere is found, the algorithm should be restarted
with the radius equal to $\|\y-\H\x_{in}\|$. The expected complexity results derived 
in Section \ref{expected_complexity} are exact for the Fincke-Pohst search strategy and can 
be viewed as an upper bound on the expected complexity of the Schnorr-Euchner scheme
with radius update.

{\em Remark 2:} The pseudo-code of the algorithm shown in Table~I assumes 
an alphabet having unit spacing, which can be generalized in a straightforward manner. 
For non-negative alphabets, the algorithm can be further improved by imposing
the condition that if a node at a given level violates the sparsity constraint, then
all the remaining nodes at that level will also violate the constraint.  Details are 
omitted for brevity.

\begin{figure}[t]
\centering
\subfigure[]{
\includegraphics[width=3.1in]{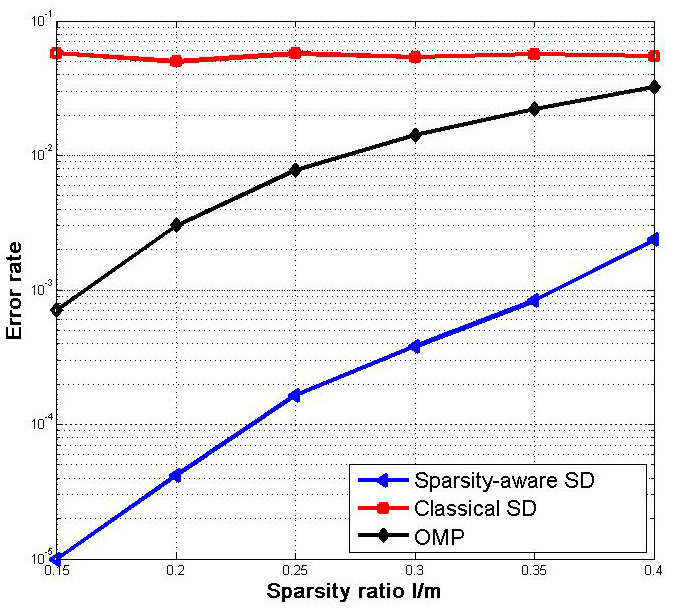}
\label{fig3a}
}
\hspace{0.25in}
\subfigure[]{
\includegraphics[width=3.1in]{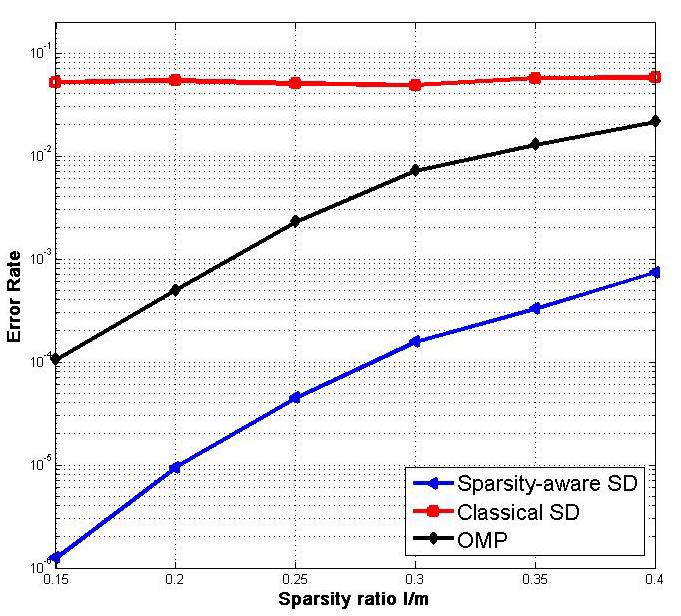}
\label{fig3b}
}
\label{fig3}
\caption[Optional caption for list of figures]{\small Error rate of sparsity-aware SD and classical SD as a function of the sparsity ratio $l/m$ in subfigure (a) and  (b) for binary $\{0,1\}$ alphabet at SNR = $10$ dB. For (a), $m=n=20$ and for (b), $m=n=40$. Performance 
of the suboptimal OMP algorithm, where the integer constraint on the entries of $\x$ is
relaxed and the solution to the relaxed problem is rounded off to nearest integer in the alphabet,  
is also shown.
}
\end{figure}

\begin{figure}
\centering
\includegraphics[width=3.4in]{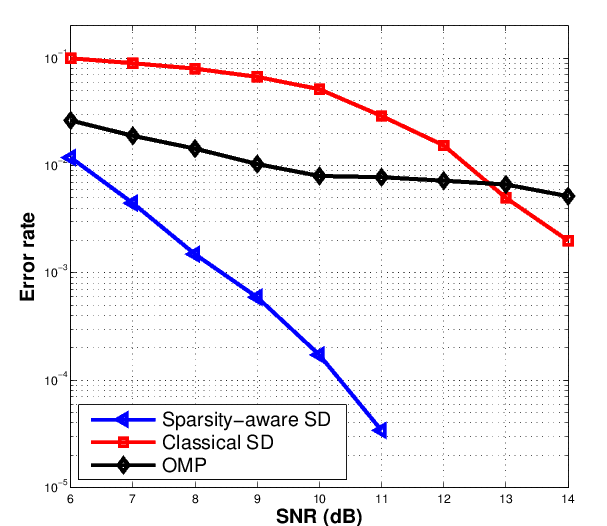}
\caption[Optional caption for list of figures]{\small Error rate of sparsity-aware SD and classical SD as a function of SNR for binary $\{0,1\}$ alphabet with $m=n=20$ and $l=5$. Performance 
of the suboptimal OMP algorithm, where the integer constraint on the entries of $\x$ is
relaxed and the solution to the relaxed problem is rounded off to nearest integer in the alphabet,  
is also shown.
\label{fig4}
}
\end{figure}

We should also point out that, in addition to helping improve computational
complexity, utilizing knowledge about sparseness of the unknown vector also
improves accuracy of the algorithm. To illustrate this, in Figure~\ref{fig3a} we
compare the error rate (i.e., the average fraction of the incorrectly inferred
components of $\x$) performance of sparsity-aware sphere decoder with 
that of the classical (sparsity-unaware) sphere decoding algorithm. In this figure, 
the error rate is plotted as a function of the sparsity ratio $l/m$. We use
a sparse binary alphabet, and simulate a system with $m=n=20$
at SNR = 10 dB. The classical sphere decoder is unaware of the sparseness of the
unknown vector (i.e., it essentially assumes $l = 20$). It is shown in the figure that the 
sparsity-aware sphere decoding algorithm performs exceptionally well compared 
to classical sphere decoder for low values of $l$. As expected, the performance 
gap between the two decoders diminishes if the unknown vector is less sparse.
For a comparison, Figure~\ref{fig3a} also shows performance of the method where the
relaxed version of the integer least-squares problem is solved via orthogonal
matching pursuit (OMP) and the result is then rounded to the nearest integer in
the symbol space. It is worthwhile mentioning that the OMP method is sub-optimal 
since there are no guarantees that it will find the closest lattice point. Therefore, its 
performance (as well as performances of other sub-optimal schemes such as 
LASSO) is generally inferior to that of sparsity-aware sphere decoder, as illustrated 
in Figure~\ref{fig3a} - Figure~\ref{fig3b}.
Figure~\ref{fig3a} shows the error rates as the function of the relative sparsity level 
$l/m$ for $m=n=20$. Figure~\ref{fig3b} illustrates the same but for $m=n=40$.
Performances of the considered algorithms in Figure~\ref{fig3a} and Figure~\ref{fig3b}
exhibit the same trends -- i.e., the performance gap between sparsity-aware SD and
classical SD widens as the sparsity ratio $l/m$ reduces. 
Figure~\ref{fig4} 
shows the error rate performance of sparsity-aware and classical sphere decoder, along with OMP,
as a function of SNR, while $l$ is fixed at $5$. In this figure too, sparsity-aware sphere decoder is seen to 
dramatically outperform its classical counterpart as well as OMP.

\section{Expected Complexity of Sparsity-Aware Sphere Decoding}
\label{expected_complexity}

As elaborated in \cite{vikalo1}, computational complexity of the sphere decoding algorithm 
depends on the effective size of the search tree, i.e., the number of nodes in the tree visited 
by the algorithm during the search for the optimal solution to the integer least-squares problem. 
If both $\H$ and $\boldsymbol\nu$ are random variables, so is the number of tree nodes
visited by the algorithm. Therefore, it is meaningful to treat the complexity as a random
variable and characterize it via a distribution or its moments. This motivates the study of
the expected complexity of the sparsity-aware sphere decoding algorithm presented next.


Assume that $\boldsymbol\nu = \y - {\bf H} \x$ (the perturbation noise) consists of independent, identically distributed
entries having Gaussian distribution $\mathcal{N}(0,\sigma^2)$. Clearly, $\frac{1}{\sigma^2}\|\boldsymbol\nu\|^2 $ 
is a random variable following $\chi_n^2$ distribution. As argued in \cite{vikalo1}, if the radius
of the search sphere is chosen as $d^2 = \alpha n \sigma^2$, where $\alpha$ is such that
$\gamma(\frac{\alpha n}{2}, \frac{n}{2}) = 1-\epsilon$\footnote{$\gamma(a,b)$ denotes the 
normalized incomplete gamma function.} for small $\epsilon > 0$, then with high probability the sphere decoding 
algorithm will visit at least one lattice point. Moreover, with such a choice of the radius, 
the probability that an arbitrary lattice point $\H{\bf x}_a$ belongs to a


sphere $\mathcal{S}_d$ 
of radius $d$ centered at $\bf y$ is 
\begin{eqnarray}
\mathcal{P}({\bf x}_a \in \mathcal{S}_d) &=& \mathcal{P}\left( \|\y-\H\x_a\|_2^2 \leq d^2 \right) \nonumber \\ 
&=& \gamma\left( \frac{d^2}{2(\sigma^2 +\|{\bf x}_a - {\bf x}_t\|^2)}, \frac{n}{2}\right) ,
 \end{eqnarray}
where ${\bf x}_t$ denotes the true value of the unknown vector $\bf x$. 

Let $N_k$ be the random variable denoting the number of points in the tree visited by
the sparsity-aware sphere decoding algorithm searching for an $l$-sparse solution to
(\ref{firsteq}). 
It is easy
to see that
\begin{align}
& \mathbb{E}\left[ N_k | ~\x_t,d^2 \right]  
 = \sum_{\eta}^{} \sum_{\substack{\x_a^k ~:~ \x_a^k \in \mathcal{D}_L^k, \\ 
\|\x_t^k-\x_a^k\|^2=\eta \\ \|\x_a^k\|_0 \leq l}}^{} 
\gamma\Big(\frac{d^2}{2(\sigma^2+\eta)}, \frac{n-m+k}{2}\Big), &
\label{sparse_exp_pts}
\end{align}
where $\x^k$ denotes the vector consisting of the last $k$-entries of $\x$, i.e., 
$\x^k = [x_{m-k+1} ~ \cdots ~ x_m]$. Note that the inner sum in 
(\ref{sparse_exp_pts}) is only over those $\x_a^k$ that satisfy the sparsity constraint.
Therefore, the expected number of points visited by our sparsity-aware sphere
decoder, averaged over all permissible ${\bf x}_t$, is given by
\begin{eqnarray}
\mathbb{E} \left[ N_k| ~ d^2\right] &=& 
\mathbb{E} \left[ \mathbb{E}\left[ N_k|~ \x_t,d^2 \right]\right]  \label{outer_expectation}\\
&=& \sum_{\substack{\x_t^k: \x_t^k \in 
\mathcal{D}_L^k \\ \|\x_t^k\|_0 \leq l}}^{}~ p(\x_t^k)~\mathbb{E}\left[ N_k | ~\x_t,d^2 \right] 
\label{prob_xt}
\end{eqnarray}
where the outer expectation in (\ref{outer_expectation}) is evaluated with respect to
$\x_t$, and $p(\x_t^k)$ in (\ref{prob_xt}) denotes the probability that $\x_t^k$ 
is the true value of $\x^k$. Note that without the sparsity constraint on $\x_t$, it would
hold that $p(\x_t^k) = 1/L^k, ~ \forall k$, where $L$ denotes the alphabet size.    
However, due to the sparsity constraint, $p(\x_t^k)$ is not uniform. As an illustration, 
consider a simple example of the $2$-dimensional $1$-sparse set on the binary
$\{0,1\}$ alphabet comprising of vectors $\{(0,0),(0,1),(1,0)\}$. If all $x_t$ are equally
likely, it is easy to see that the probability of $\x_t^1$ is given by 
$p(\x_t^1 = 1) = 1/3 = 1- p(\x_t^1 = 0)$. For a general $l$-sparse $m$-dimensional 
constellation, the probability distribution $p(\x_t^k)$ can be obtained by 
enumerating all $k$-dimensional vectors in the set for $k=1, 2, \cdots, m$.

Having determined expected number of lattice points visited by the sparsity-aware
sphere decoding algorithm, the average complexity can be expressed as
\begin{eqnarray}
& C(m,d^2) = \sum\limits_{k=1}^{m}~f(k) ~\mathbb{E} \left[ N_k |~ d^2\right], & 
\label{complexity_expression}
\end{eqnarray}
where $f(k)$ denotes the number of operations performed by the algorithm in the $k^{th}$ 
dimension. 

The main challenge in evaluating (\ref{sparse_exp_pts}) is to find an efficient enumeration of the 
symbol space, i.e., to determine the number of $l$-sparse vectors ${\bf x}_a^k$ such that 
$\|{\bf x}_t^k-{\bf x}_a^k\|^2 = \eta$ for a given $\x_t^k$. While this enumeration appears to 
be difficult in general, it can be found in a closed form for some of the most commonly 
encountered alphabets in sparse integer least square problems: the binary $\{0,1\}$ alphabet 
(relevant to applications in \cite{collu1}, \cite{collu2} and \cite{mitra}) and the ternary
$\{-1,0,1\}$ alphabet (relevant to application in \cite{microarrays}).  In the rest of this section, 
we provide closed form expressions of the expected complexity of sparsity-aware sphere
decoder for these alphabets. 

\subsection{Binary Alphabet \{0,1\}}

Recall that computing (\ref{sparse_exp_pts}) requires enumeration of the sparse symbol space, 
i.e., counting $l$-sparse vectors $\x_a^k \in \mathcal{D}_L^k$ satisfying 
$\|\x_a^k-\x_t^k\|^2 = \eta$ for a given $l$-sparse vector $\x_t^k \in \mathcal{D}_L^k$ 
and $\eta$. Note that, for the binary alphabet, condition $\|\x_t^k-\x_a^k\|_2^2 = \eta$ 
is equivalent to $\|\x_t^k-\x_a^k\|_0 = \eta$. Let $\|\x_t^k\|_0 = k_1$, and denote the $i^{th}$ 
entry of $\x^k$ as $\x^k(i), ~ i = 1, \cdots, k$. Furthermore, let $\|\x_t\|_0 = k_3$.

{\it Lemma 1:} Given $k_1 = \|\x_t^k\|_0$ and $\eta$, the number of $k$-dimensional lattice points 
$\x_a^k$ with $\|\x_a^k\|_0 = k_2$ such that $\|x_a^k-x_t^k\|^2 = \eta$ is given by 
\begin{equation}
g(k_1,k_2,k,\eta) = \binom{k_1}{p} \binom{k-k_1}{q}, 
\label{gdefn}
\end{equation}
where
\begin{eqnarray} 
p = \frac{1}{2}(\eta-(k_2-k_1)),~ q = k-k_2-p,  & \mbox{if} & k_1 < k_2  \nonumber \\
q = \frac{1}{2}(\eta-(k_1-k_2)),  ~p = k_2-q,  & \mbox{if} & k_1 \geq k_2 \nonumber
\end{eqnarray}
\begin{proof}
See Appendix A.
\end{proof}
\vspace{2mm}

Note that for a given $\x_t^k$ and $k_2$, the possible values of $\eta$ belong to the
set $\mathcal{S} = \{|k_1-k_2|, |k_1-k_2|+2, \cdots, \min{(k_1+k_2,k)}\}$. 
Then, (\ref{sparse_exp_pts}) can be written as 
\begin{align}
& \mathbb{E}[N_k |~ \x_t, d^2]  
 = \sum_{k_2=0}^{\min(k,l)} 
\sum\limits_{\eta \in \mathcal{S}}^{}\gamma\left(\frac{d^2}{2(\sigma^2+\eta)}, \frac{n-m+k}
{2}\right)g(k_1,k_2, k,\eta). & \label{mid_exp}
\end{align}

Finally, outer expectation in (\ref{outer_expectation}) is enumerated as follows. 
Total number of sparse binary vectors $\x_t^k$ for a given $k$ can be parameterized 
by $k_1$ and $k_3$. For $\x_t$ to be $l$-sparse, $0 \leq k_3 \leq l$ 
and for each $k_3$, it should hold that $(k_3-(m-k))_{+} \leq  k_1 \leq \min{(k,k_3)}$. 
The number of $\x_t$ vectors for each pair of values of $k_3$ and $k_1$ is then 
given by $\binom{k}{k_1}\binom{m-k}{k_3-k_1}$. Therefore, the total number of all 
possible $l$-sparse vectors $\x_t$ is given as $N  = \sum\limits_{k_3, k_1}^{}  
\binom{k}{k_1}\binom{m-k}{k_3-k_1}$ and  (\ref{prob_xt}) can be written in terms of 
$k_3$ and $k_1$ as 
\begin{align}
&\mathbb{E} \left[ N_k| ~ d^2\right] = 
\frac{1}{N} \sum\limits_{k_3 ,k_1}^{}~ \binom{k}{k_1}\binom{m-k}{k_3-k_1} ~ \mathbb{E} \left[ N_k | ~ \x_t,d^2 \right] & 
\label{new_prob_xt}
\end{align}
where the summations are evaluated over respective ranges as described above. 

{\it Theorem 1}: 
The expected number of tree points examined by the sparsity-aware
sphere decoding algorithm while solving (\ref{firsteq}) over a binary alphabet is
\begin{align}
&\mathbb{E}\left[ N_k |d^2\right] = \frac{1}{N} \sum\limits_{k_3 = 0}^{l}~~~ \sum\limits_{k_1  = {(k_3 - (m-k))}_{+}}^{\min{(k,k_3)}} \binom{k}{k_1}\binom{m-k}{k_3-k_1}  
 \Big[ \sum_{k_2=0}^{\min(k,l)} 
\sum\limits_{\eta \in \mathcal{S}}^{}\gamma\left(\frac{d^2}{2(\sigma^2+\eta)}, \frac{n-m+k}
{2}\right)g(k_1,k_2, k,\eta)\Big], &
\label{case3}
\end{align}
where $d$ denotes the search radius and $g(k_1,k_2, k,\eta)$ is defined in (\ref{gdefn}). 

\noindent
\begin{proof} The proof follows from (\ref{sparse_exp_pts}) and Lemma 1.

\end{proof}

%

{\it Remark:} Note that (\ref{case3}) has important implications on the worst-case complexity 
of the sparsity-aware sphere decoding algorithm. Assume, for the sake of argument, that the 
radius $d$ of the sphere is sufficiently large to cover the entire lattice. In the absence of the
radius constraint, pruning of the search tree happens only when the sparsity constraint is
violated, and hence (\ref{mid_exp}) reduces to
\begin{eqnarray}
\mathbb{E}\left[ N_k |~\x_t,d^2\right] &=&
 \sum_{k_2=0}^{\min(k,\ell)} 
\sum\limits_{\eta \in \mathcal{S}}^{}g(k_1,k_2, k,\eta)   \nonumber \\
&=& \sum_{k_2=0}^{\min(k,\ell)} \binom{k}{k_2}.
\label{case4}
\end{eqnarray}
Expression (\ref{case4}) essentially represents the worst-case scenario where a brute-force 
search over all $l$-sparse signals is required, leading to a complexity exponential in $k$. 
For $k < l$, this is clearly the case. For $k \geq l$, the partial sum of binomial coefficients 
in (\ref{case4}) cannot be computed in a closed form but its various approximations and 
bounds exists in literature. For example, it has been shown in \cite{worsch} that the partial 
sum of binomial coefficients in (\ref{case4}) grows exponentially in $k$ if $l = k/a$ for a 
constant $a \geq 2$. A corollary to this result states that if $a$ is an unbounded, 
monotonically increasing function, the sum in (\ref{case4}) does not grow exponentially in 
$k$. Therefore, unless the fraction of the non-zero components of $\x$ becomes vanishingly 
small as the dimension of $\x$ grows, the worst-case complexity of the sparsity-aware 
sphere decoding algorithm is exponential in $m$ (the length of $\x$).


Also note that if $\x_t$ is $l$-sparse but the sparsity constraint is not invoked at the decoder 
(i.e., the decoder is sparsity-unaware), then (\ref{mid_exp}) is reduced to
\begin{align}
&\mathbb{E}\left[ N_k |~\x_t,d^2\right] =
 \sum_{k_2=0}^{k} 
\sum\limits_{\eta \in \mathcal{S}}^{}\gamma\left( \frac{d^2}{2(\sigma^2+\eta)}, \frac{n-m+k}{2}\right)g(k_1,k_2, k,\eta). &
\label{case5}
\end{align}
Clearly, (\ref{case5}) is an upper bound for (\ref{mid_exp}), and hence exploiting sparsity 
information enables reduction of the expected complexity.

\subsection{Ternary Alphabet \{-1,0,1\}}

Define the support sets $S_j(\x_t^k)$ of $\x_t^k$ for $j \in \{-1, 0, 1\}$  
and let $a = |S_1(\x_t^k)|$. Unlike the binary case, here we evaluate (\ref{sparse_exp_pts}) by 
reversing the order of summation, i.e., by enumerating all possible $l$-sparse vectors 
$\x_a^k$ and then summing over all $\eta$ such that 
$\|\x_t^k  - \x_a^k\|_2^2 = \eta$. To this end, let us introduce \vspace{-2mm}
\begin{eqnarray}
p_{i,j} = \sum\limits_{r \in S_i(\x_t^k) }^{}\mathcal{I}_{\{\x_a^k(r)=j\}}, \;\;\; i,j \in \{-1, 0, 1\}.
\end{eqnarray}
In words, $p_{i,j}$ denote the number of symbols $j$ in $\x_a^k$ in the positions where 
$\x_t^k$ has symbol $i$. It is easy to see that $\sum_{i}\sum_{j} p_{i,j} = k$ and, furthermore, 
\begin{eqnarray}
\sum\limits_{j}^{}p_{i,j} = \begin{cases} a, & i=1 \\
k_1-a, & i = -1 \\
k-k_1, & i = 0.
\end{cases} &&
\label{p_rel}
\end{eqnarray}
where $k_1 = \|\x_t^k\|_0$.
Given $p_{i,j}$, $\eta$ can be written as 
\begin{small}
\begin{eqnarray} 
\eta &=& \sum\limits_{i,j}^{} p_{i,j} |i-j|^2. \label{eta_eqn}
\end{eqnarray}\end{small}

{\it Lemma 2}: Given $k_1 = \|\x_t^k\|_0$ and $\eta$, the number of $k$-dimensional lattice 
points $\x_a^k$ with $\|\x_a^k\|_0 = k_2$ such that $\|x_a^k-x_t^k\|^2 = \eta$ is given by 
\begin{small}
\begin{align}
 & g_1(k_1, k_2,k,p_{ij}) = \binom{a}{p_{1,0}}\binom{a-p_{1,0}}{p_{1,-1}} \binom{k_1-a}{p_{-1,0}}\binom{k_1-a-p_{-1,0}}{p_{-1,1}} 
\binom{k-k_1}{k_2-(k_1-p_{-1,0}-p_{1,0})}2^{k_2-(k_1-p_{-1,0}-p_{1,0})}, & \label{ternary_case1}
\end{align} \end{small} \vspace{-4mm}
for $k_1 \leq k_2$ and 
\begin{small}
\begin{align}
&g_2(k_1, k_2,k, p_{ij}) = \binom{k-k_1}{p_{0,1}+ p_{0,-1}}\binom{a}{p_{1,-1}} \binom{a-p_{1,-1}}{p_{1,1}}  \binom{k_1-a}{p_{-1,1}} 
 \binom{k_1-a- p_{-1,1}}{k_2-(p_{0,1}+p_{0,-1}+p_{1,-1}+p_{1,1}+p_{-1,1})}2^{p_{0,1}+p_{0,-1}}, & \label{ternary_case2}
\end{align} 
\end{small}
for $k_1 > k_2$.

\begin{proof}
See Appendix B.
\end{proof}
Similar to the binary case, it can be shown that the total number of $l$-sparse vectors 
$\x_t$ is given by $\binom{k}{a}\binom{k-a}{k_1-a}\binom{m-k}{k_3-k_1}2^{k_3-k_1}$, 
where $k_3 = \|\x_t\|_0$, $0 \leq k_3 \leq l$, ${(l-(m-k))}_{+} \leq k_1 \leq  \min{(k, k_3)}$ 
and $0 \leq a \leq k_1$. Therefore, (\ref{prob_xt}) can be written in this case as 
\begin{align*}
& \mathbb{E}[N_k ~|~ d^2] =  \frac{1}{N} \sum\limits_{k_3, k_1, a}^{} \binom{k}{a}\binom{k-a}{k_1-a}\binom{m-k}{k_3-k_1}2^{k_3-k_1}
 ~ \mathbb{E}[N_k | ~ \x_t^k, d^2], &
\end{align*}
where $N$ is the number of all $l$-sparse vectors $\x_t$, given by $N = \sum\limits_{k_3, k_1, a}^{} \binom{k}{a}\binom{k-a}{k_1-a}\binom{m-k}{k_3-k_1}2^{k_3-k_1}$, and the summations are evaluated over respective ranges as described above. 

{\it Theorem 2}: 
The expected number of tree points examined by the sparsity-aware
sphere decoding algorithm while solving (\ref{firsteq}) over a ternary alphabet is
\begin{align}
& \mathbb{E}[N_k |~  d^2] = 
 \frac{1}{N} \sum\limits_{k_3 = 0}^{l} ~~
\sum\limits_{k_1 = {(l-(m-k))}_{+}}^{\min{(k,k_3)}} \sum\limits_{a = 0}^{k_1} \binom{k}{a}\binom{k-a}{k_1-a}\binom{m-k}{k_3-k_1}2^{k_3-k_1} & \nonumber \\
& \times \Bigg[  \sum\limits_{k_2=0}^{k_1-1} ~~\sum\limits_{p_{i,j} ~:~ k_1 > k_2}^{} \gamma\left(\frac{d^2}{2(\sigma^2+\eta)},\frac{n-m+k}{2}\right) 
g_2(k_1,k_2,k,p_{i,j})  \Big. & \nonumber  \\
& \Big. + \sum\limits_{k_2=k_1}^{\min{(k,l)}}~~ \sum\limits_{p_{i,j} ~:~ k_1 \leq k_2}^{}
~ \gamma\left(\frac{d^2}{2(\sigma^2+\eta)},\frac{n-m+k}{2}\right) g_1(k_1,k_2,k,p_{i,j}) \Bigg] &
\end{align}
where the value of $\eta$ is given by (\ref{eta_eqn}).
\noindent

\begin{proof} The proof follows from (\ref{sparse_exp_pts}) and Lemma 2.

\end{proof}

\subsection{Results}

\begin{figure}[!t]
\begin{center}	
	\includegraphics[height=3in]{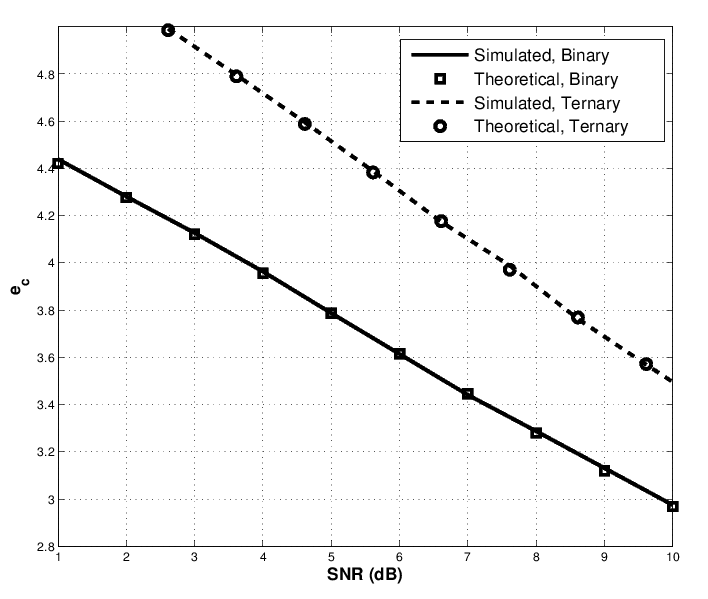}	
	\vspace{-3mm}
	\caption{\small Complexity exponent as a function of SNR for $m=n=20$, $l=5$ for binary $\{0,1\}$ and ternary $\{-1,0,1\}$ alphabets.}
			\label{binary_ternary_complexity}
	\end{center}
    \end{figure}

It is useful to define the complexity exponent, $e_c = \log{C(m,d^2)}/ \log{m}$, 
as a measure of complexity of the algorithm.
To validate the derived theoretical expressions for the expected complexity of the sparsity-aware 
sphere decoding algorithm, we compare them with empirically evaluated complexity exponents.
The results are shown in Fig.~\ref{binary_ternary_complexity}. Here, parameters of the simulated
system are $m=n=20$ and $l=5$. The empirical expected complexity is obtained by averaging
over $1000$ Monte Carlo runs. As can be seen in in Fig.~\ref{binary_ternary_complexity}, the
theoretical expressions derived in this section exactly match empirical results and are hence
corroborated. 

    \begin{figure}[!t]
 \begin{center}
	\includegraphics[height=3in]{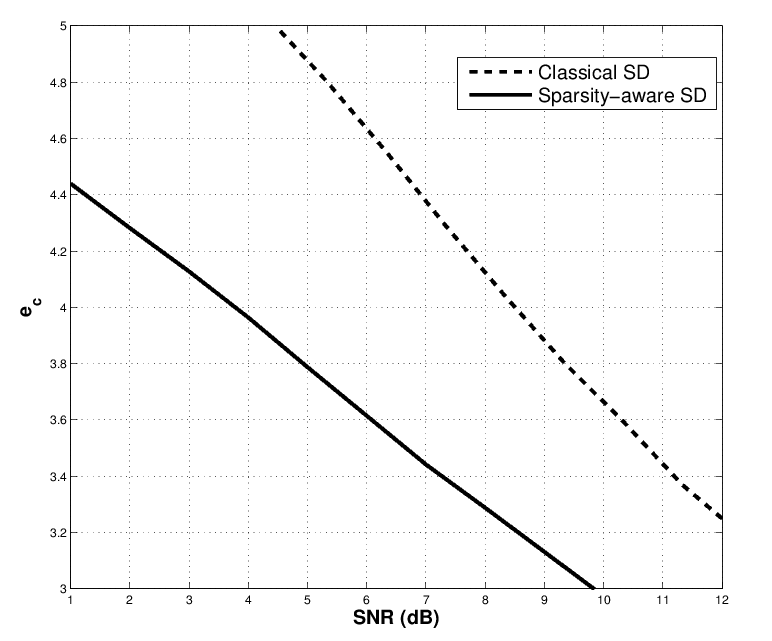} 
	\caption{\small Comparison of the complexity exponent of sparsity-aware sphere decoder and 
classical (sparsity-unaware) sphere decoder for $m=n=20$, $l=5$ and binary $\{0,1\}$ alphabet.}
	\label{binary_classical_complexity}
    \end{center}\end{figure}

Fig.~\ref{binary_classical_complexity} compares the complexity exponent of sparsity-aware 
sphere decoder with that of the classical (sparsity-unaware) sphere decoding algorithm for 
$m=n=20$, $l=5$ and binary $\{0,1\}$ alphabet. Similar to Fig. \ref{fig3a} and \ref{fig3b}, classical sphere 
decoder assumes the unknown vector to be non-sparse. As can be seen from the plot, the 
complexity of sparsity-aware sphere decoder is less than that of classical decoder, with the 
gap being more significant at lower values of signal-to-noise ratios (SNRs) and decreasing as 
SNR increases. This is expected because, for low SNRs, radius of the sphere is large and the 
complexity of sparsity-aware sphere decoder approaches that of performing an exhaustive 
search for the closest point. Since the cardinality of the sparse set in this example is 
significantly smaller than that of the full lattice, the complexity gap between two algorithms 
is pronounced. However, for high SNRs, the sphere often contains very few lattice point
(occasionally only one), and hence the difference between the complexity exponents of the 
two algorithms reduces. Also note that the complexity of sparsity-aware sphere decoder 
decreases with $l$ since the sparsity constraint further reduces the number of points that
will be in a sphere of a given radius.

\section{Variance Analysis}

In this section, we characterize the second-order moment of the complexity of 
sparsity-aware sphere decoder. The following analysis is restricted to the binary 
alphabet, but it can be extended to more general alphabets in a relatively straightforward manner. It has been 
shown in \cite{vikalo2} that the variance of the complexity of the sphere decoding 
algorithm is given by
\begin{align}
& \sigma_{\mbox{\small sparse}}^2  =  
 \sum\limits_{k=1}^{m} \sum\limits_{l=1}^{m} \left( \mathbb{E}[N_kN_l] 
- ~\mathbb{E}[N_k] \mathbb{E}[N_l]\right) f(k)f(l), & \label{vareq} 
\end{align}
where $N_k$ and $f(k)$ have same meanings as in Section~III.
Note that (\ref{vareq}) applies to the sparsity-aware sphere decoder as well, but
$\mathbb{E}[N_k]$, $\mathbb{E}[N_l]$, and $\mathbb{E}[N_k N_l]$ differ from 
those for the classical (sparsity-unaware) algorithm.
In Section \ref{expected_complexity}, we found an expression for the expected number
of points $\mathbb{E}[N_k]$ visited by the sparsity-aware sphere decoding algorithm. 
Therefore, to evaluate (\ref{vareq}), what remains to be determined is $\mathbb{E}[N_k N_l]$, 
i.e., the correlation between the number of pairs of points of dimensions $k$ and $l$ lying 
inside a sphere of radius $d$. Let each of these points be $l^{\prime}$-sparse.

Given $\H = \Q\R$, and using (\ref{sphrcons}), the correlation between $N_k$ and $N_l$ can 
be found as \cite{vikalo2}
\begin{align}
\mathbb{E}[N_kN_l]  = \sum\limits_{\x_B \in \mathcal{D}_L^k}^{} ~\sum\limits_{\x_C \in \mathcal{D}_L^l}^{}~\mathcal{P}\left(t_b \leq d^2,t_c \leq d^2 \right),  && \label{enumeration}
\end{align}
where $t_b = \|\z^k-\R(k,k)\x_B\|^2$, $t_c = \|\z^l-\R(l,l)\x_C\|^2$,
$\x_B$ and $\x_C$ are arbitrary $k$ and $l$ dimensional $l^{\prime}$-sparse vectors having entries from
${\cal D}_L$, and $\R(p,p)$ is given by the following partition of $\R$,
\begin{eqnarray}
& \R =
\left[\begin{array}{cc}
\R(m-p,m-p) & \R(m-p,p) \\
{\bf 0}_{p \times (m-p)} & \R(p,p) 
\end{array} \right]&
\end{eqnarray}
for $p=1, \ldots, m$.
Let $\x_t^k$ be the $k$-dimensional sub-vector of the true solution, and define 
$\x_b = \x_t^k-\x_B$ and 
$\x_c = \x_t^l - \x_C$. Depending on whether $\x_b= \x_c^k$ or not, 
the summand in (\ref{enumeration}) is given by \cite{vikalo2}:

\begin{enumerate}

\item If $x_b = x_c^{k}$,
\begin{equation}
\mathcal{P} \left(t_b \le d^2, t_c \le d^2 \right) =
\gamma \left(\frac{d^2}{2(\sigma^2+\|x_c\|^2)},\frac{l}{2}\right)
\label{probvar1}
\end{equation}

\item If $x_b \ne x_c^{k}$,
\begin{equation}
\mathcal{P}(t_b \le d^2, t_c \le d^2) = \int_{t_b=0}^{d^2} \int_{t_c=0}^{d^2}
\phi(t_b,t_c) dt_b dt_c,
\label{probvar2}
\end{equation}
where $\phi(t_b,t_c)$ is given by (\ref{charfun}),
\begin{figure*}[!b]
\begin{equation}
\phi(t_b,t_c) = \frac{1}{4\pi^2}
\int_{-\infty}^\infty \int_{-\infty}^\infty
\frac{d\omega_b d\omega_c e^{-j\omega_b t_b - j \omega_c t_c}}
{\Delta^{k/2} \left[ \left(\frac{a_{bb}}{\Delta}-2j\omega_c\right)
\left(\frac{a_{cc}}{\Delta}-2j\omega_b\right)
-\frac{a_{bc}^2}{\Delta^2}\right]^{k/2}}
\frac{\left(1-2j\omega_c(\sigma^2+\|x_c^{k}\|^2)\right)^{k/2}}
{\left(1-2j\omega_c(\sigma^2+\|x_c\|^2)\right)^{l/2}},
\label{charfun}
\end{equation}
\end{figure*}
and where $a_{bb} = \|x_b\|^2+\sigma^2$, $a_{cc} = \|x_c^{k}\|^2+\sigma^2$, 
$a_{bc} = x_b^*x_c^{k}+\sigma^2$, and $\Delta = a_{bb}a_{cc}-a_{bc}^2$.
\end{enumerate}

The number of pairs of points ($\x_B, \x_C$) over which the summation in 
(\ref{enumeration}) is evaluated depends on the specific symbol alphabet. 
Here we outline how to enumerate the total number of such pairs for the 
binary alphabet $\{0,1\}$. Let us assume, without a loss of generality, that 
$k \leq l$. As shown in \cite{vikalo2}, $p\left(t_b \leq d^2,t_c \leq d^2 \right)$
is a function of $\|\x_b\|^2$, $\|\x_c^k\|^2$, $\|\x_c\|^2$ and
$\x_b^{\star}\x_c^k$. Therefore, we can evaluate (\ref{enumeration}) by 
counting the number of all possible solutions ($\x_b, \x_c$) to the system of 
equations
\begin{eqnarray}
\|\x_b\|^2 = \b, ~ \|\x_c\|^2 = \e, ~ \|\x_c^k\|^2 = \g, ~~\mbox{and}~~ 
\x_b^{\star}\x_c^k = \d, \label{seteq}
\end{eqnarray}
where 
$\b, \e,\g$ and $\d$ are integer numbers satisfying the constraints imposed 
by the dimensions $k$ and $l$. Unlike the scenario studied in \cite{vikalo2}, 
the space of permissible solutions to our problem is not isotropic (due to
sparsity constraint). Note that since $\x_B$ and $\x_C$ belong to $\{0,1\}$ 
alphabet, $\x_b$ and $\x_c$ are $k$ and $l$-dimensional vectors with entries 
from the set $\{-1,0,1\}$. Moreover, each of these vectors is $2l^{\prime}$-sparse. 
For the binary alphabet, $\ell_0$ norm is equivalent to $\ell_2$ norm. Therefore, 
the range of values that $\beta = \|\x_b\|_0$ can take is 
$\mathcal{S}(\b)$= $\{0,1,\ldots,\min{(k,2l^{\prime})}\}$ and, similarly, the range of 
values that $\e = \|\x_c\|_0$  can take is  $\mathcal{S}({\e})$ = 
$\{0,1,\ldots,\min{(l,2l^{\prime})}\}$. It is straightforward to see that for a given value 
of $\e$, the range of $\g$ is defined by
$\mathcal{S}({\g}|\e)$ = $\{(\e-(l-k))_{+}, \ldots, \min{(\e,k)}\}$, where 
$(a)_+ = a$ if $a \geq 0$, 0  otherwise. Now, non-zero entries ($-1$ and $1$)
in $\x_c^k$ and $\x_c$ -- let us denote their number by $\g$ and $\eta$, respectively -- can be arranged in $\binom{l-k}{\eta-\g}\binom{k}{\g}$ ways.
Therefore, the number of possible $l$-dimensional vectors $\x_c$ with 
$\|\x_c\|_0 = \e$ and $\|\x_c^k\|_0 = \g$ is given by 
\begin{eqnarray}
& \sum\limits_{\e \in 
\mathcal{S}(\e)} \sum\limits_{\g \in \mathcal{S}(\g|\e) } \binom{l-k}{\e-\g}  \binom{k}{\g} 2^{\e}. & \label{xc_number}
\end{eqnarray}
For any given $\x_c^k$ and $\d = \x_b^{\star}\x_c^k$, we proceed by finding 
the number of possible pairs ($\x_b, \x_c$) satisfying the conditions (\ref{seteq}). 
A close inspection of the definitions of $\x_b$ and $\x_c$ reveals that $\d$ 
corresponds to the number of positions (out of $k$) where the entries of $\x_b$ 
and $\x_c^k$ can take values $(1,1)$ or $(-1,-1)$.  Let us define 
\begin{eqnarray}
& u = \sum\limits_{t \in \mathcal{S}_1(\x_c^k)}{} \mathcal{I}_{\{\x_b(t)=1\}}~ \mbox {and}~~ v = \sum\limits_{t \in \mathcal{S}_{-1}(\x_c^k)}{} \mathcal{I}_{\{\x_b(t)=-1\}}, & \nonumber 
\end{eqnarray}
where $\mathcal{S}_1(\x_c^k)$ and $\mathcal{S}_{-1}(\x_c^k)$ denote sets of indices 
of entries in $\x_c^k$ valued $1$ and $-1$, respectively. Clearly, we have $\d = u+v$.
Let $a =|\mathcal{S}_1(\x_c^k)|$. Now, $u$ $1$'s and $v$ $-1$'s in $\x_b$ can be 
arranged in $\binom{a}{u}\binom{\g-a}{v}$ ways, while the remaining $\b-(u+v)$ 
non-zero entries of $\x_b$ can be arranged in $k-\g$ zero-valued positions of
$\x_c^k$ in $\binom{k-\g}{\b-(u+v)}2^{\b-(u+v)}$ ways. Therefore, the number of 
possible $k$-dimensional vectors $\x_b$ satisfying (\ref{seteq}) for a given 
$\x_c^k$ is $\binom{a}{u}\binom{\g-a}{v}\binom{k-\g}{\b-(u+v)}2^{\b-(u+v)}$. 

All that remains to be done now is to define the admissible range of values of $u$ 
and $v$. It can be easily shown that the set for $u$ is given by 
$\mathcal{S}(u|a,\b) = \{(\b-(k-a))_{+}, \ldots, \min{(\b,a)}\}$.
For each $u \in \mathcal{S}(u|a,\b)$, $v$ can take values from the set 
$\mathcal{S}(v|u,a,\b,\g) = \{(\b-(u+k-\g))_{+}, \ldots, \min{(\g-a,\b-u)}\}$. 

Using above results and (\ref{xc_number}), the total number of possible solutions to the set of equations 
(\ref{seteq}) is given by
\begin{align}
&g(\b,\e,\g,a,u,v) =   
  \binom{l-k}{\e-\g}  \binom{k}{\g}  
\binom{\g}{a}  \binom{a}{u}\binom{\g-a}{v}\binom{k-\g}{\b-\d}2^{\e-\g +\b-\d}, & \label{count}
\end{align}

\noindent
where $\d = u+v$. For $l <k$, the set of equations 
(\ref{seteq}) are changed to
\begin{eqnarray}
\|\x_c\|^2 = \b, ~ \|\x_b\|^2 = \e, ~ \|\x_b^l\|^2 = \g ~~\mbox{and}~~ \x_c^{\star}\x_b^l = \d, \label{seteq2}
\end{eqnarray}
and the roles of $l$ and $k$ are reversed in (\ref{count}). 

Finally, using (\ref{enumeration}), the correlation between $N_k$ and $N_l$ is given by
\begin{align}
&\mathbb{E}[N_kN_l ]  
 = \sum\limits_{\e,\g,\b,a, p,q}^{}g(\b,\e,\g,a,u,v)~ \mathcal{P}(t_b \leq d^2, t_c \leq d^2), & 
\label{jointexpect}
\end{align}
where the summation in (\ref{jointexpect}) is taken over respective sets for the variables as 
given by $\mathcal{S}(\e), \mathcal{S}(\g|\e), \ldots,$ $ \mathcal{S}(v|u,a,\b,\g)$, etc. 
Substituting (\ref{jointexpect}) and (\ref{prob_xt}) in (\ref{vareq}), we 
obtain the expression for the variance of the complexity of the sparsity-aware sphere 
decoding algorithm for binary alphabet.

\section{Speeding-up sparsity-aware sphere decoder using lower bounds on the objective 
function}
\label{speedup}
In this section, we present a method for reducing complexity of the sparsity-aware sphere 
decoding algorithm by performing additional pruning of the nodes from the search tree. In 
particular, motivated by the lower bounding idea of \cite{Hinfty} and observing that the
solution to the $\ell_1$-norm regularized relaxed version of the optimization problem 
(\ref{firsteq}) can be obtained relatively inexpensively, we formulate a technique capable 
of significant reduction of the total number of tree points that are visited during the 
sparsity-aware sphere decoder search.

For convenience, let ${\bf z}_{a:b}$ denote the vector comprising entries of ${\bf z}$ 
indexed $a$ through 
$b$, $a \leq b$. Similarly, let $\A_{a:c,b:d}$ denote the sub-matrix of $\A$ collecting
entries with indices $(a,b)$ (in the upper-left corner) through $(c,d)$ (in the bottom-right 
corner), $a \leq c,b \leq d $. Let $\Q_1$, $\Q_2$, and $\R$ be the matrices
defined in Section \ref{sd_overview}. Using the triangular property of $\R$ and defining 
$\z = \Q_1^{\star}\y$, we can rewrite (\ref{sphrcons}) as 
\begin{eqnarray}
& {d}^2  \geq  \|{\bf z}_{k:m} - \R_{k:m,k:m}{\bf x}_{k:m}\|^2  
 + \|{\bf z}_{1:k-1} - \R_{1:k-1,1:k-1}{\bf x}_{1:k-1} - \R_{1:k-1,k:m}{\bf x}_{k:m} \|^2, & 
\label{lowerbnd1}	
\end{eqnarray}
for any $2 \leq k \leq m$. Note that, when the sphere decoding algorithm visits a node at
the $k^{th}$ level of the search tree, vector ${\bf x}_{k+1:m}$ is already chosen and the
only variable component in the first term on the right-hand side of (\ref{lowerbnd1}) is
${x}_{k}$. Now, if we could replace the second term on the right-hand side of
(\ref{lowerbnd1}) by its lower bound $\mathcal{L}_k({\bf z}, \R, \x_{k:m}, \tilde l_k)$, defined by
\begin{eqnarray}
  \hspace{-5mm}\mathcal{L}_k({\bf z}, \R, \x_{k:m}, \tilde l_k) =
 \underset{{\bf x} \in \mathcal{D}_L^{k-1} \atop 
 \|{\bf x}\|_0 \leq \tilde{l}_k}{\min} \|{\bf w}_{1:k-1} 
 - \R_{1:k-1,1:k-1} {\bf x}\|^2 &&   \label{lowerbnd2}
\end{eqnarray}
where $\tilde l_k = (l-\|{\bf x}_{k:m}\|_0)_+$ and
${\bf w}_{1:k-1} = {\bf z}_{1:k-1} - \R_{1:k-1,k:m}{\bf x}_{k:m}$, we would obtain a more 
rigorous constraint on ${x}_{k}$ and hence traverse the search tree more efficiently, 
pruning more points. In particular, we now may require that
\[ 
{d}^2 - \mathcal{L}_k({\bf z},\R, \x_{k:m}, \tilde l_k) \geq
\|{\bf z}_{k:m} - \R_{k:m,k:m}{\bf x}_{k:m}\|^2, 
\]
which is a more strict constraint on  ${x}_{k}$ than the one used by the original sphere decoder,
\[
 {d^{\prime}}^2 \geq \|{\bf z}_{k:m} - \R_{k:m,k:m} {\bf x}_{k:m}\|^2,
\]
leading to a search where fewer tree points are being visited by the algorithm. Since the
second term on the right-hand side of (\ref{lowerbnd1}) is replaced by its lower bound,
it is guaranteed that the optimal point will not be discarded due to having a more rigorous 
constraint.

Note that the minimization in (\ref{lowerbnd2}) is a sparse integer least squares problem 
with an adaptively updated sparsity constraint $\tilde l_k = (l-\|{\bf x}_{k:m}\|_0)_+$,
i.e., the number of allowed non-zero entries in ${\bf x}_{1:k-1}$ is adjusted according
to the number of non-zero entries in already selected ${\bf x}_{k:m}$.
Clearly, it must be satisfied that $\tilde l_2 \leq \tilde l_3 \leq \ldots \leq \tilde l_m$. 
Having $\tilde{l}_k=0$ for any $k$ implies that no more non-zero elements can be chosen
as the components of the unknown vector at levels beyond the $k^{th}$, and the remaining
entries are all valued $0$. 

The modified algorithm is presented in Table \ref{algo2}.
\begin{Huge}
\begin{table}
\begin{center}
\caption{{\small Reduced-complexity sparsity-aware sphere decoding algorithm using lower 
bounds on the objective function.}}
\begin{tabular}{l}
\hline
 Input: $Q = [Q_1 ~~ Q_2], ~ R, y, ~z = Q_1^*y,$ 
 sphere radius $d$, \\ sparsity constraint $l$. \\
\hline \\
 1. \underline{ Initialize}  $k \leftarrow m$, $d_m^2 \leftarrow d^2- \|Q_2^*y\|^2$, \\
$z_{m|m+1} \leftarrow z_m$,  $l_m \leftarrow 0$ . \\
2. \underline{ Update Interval} ~~
 $UB(x_k) \leftarrow \left \lfloor (d_k + z_{k|k+1})/R_{k,k} \right \rfloor$,  
\\ $LB(x_k) \leftarrow \left \lceil  (-d_k + z_{k|k+1})/R_{k,k} \right \rceil  $,
 $x_k \leftarrow LB(x_k)-1$. \\
3. \underline{ Update $x_k$}   ~~ $x_k \leftarrow x_k+1$.
If $x_k \leq UB(x_k),$ \\ go to  4; else go to 6. \\
4. \underline{Check Sparsity} ~~~ If $l_k+ \mathcal{I}_{\{x_k \neq 0\}} > l,~$ 
go to 3; \\ else if $k >1$, go to 5; 
else $l_k \leftarrow l_k + \mathcal{I}_{\{x_k \neq 0\}}$, and go to  7. \\
5. \underline{Lower Bound} $\tilde l_k \leftarrow l-(l_k + \mathcal{I}_{\{x_k \neq 0\}})$, and  \\
compute $\mathcal{L} = \mathcal{L}({\bf z}, R, \x_{k:m}, \tilde l_k)$. \\
If $\mathcal{L}+ 
(z_{k|k+1} - R_{k,k}x_k)^2  \leq d_k^2$,  
$l_k \leftarrow l_k + \mathcal{I}_{\{x_k \neq 0\}}$ \\
and go to 7; else go to 3. \\ 
6. \underline{Increase $k$}   ~~ $k \leftarrow k+1$. 
If  $k=m+1,$ stop; \\
else, $l_k \leftarrow l_k - \mathcal{I}_{\{x_k \neq 0\}}$ and go to 3. \\
7. \underline{Decrease $k$} If $k=1,$ go to 8; else $k \leftarrow k-1$, \\
$z_{k|k+1} \leftarrow z_k - \sum_{j=k+1}^{m}R_{k,j}x_j, $~ \\
$d_k^{2} \leftarrow d_{k+1}^{2} $
$-(z_{k+1|k+2} - R_{k+1,k+1}x_{k+1})^2,$  and go to 2. \\
8. \underline{Solution found}~~ Save $x$ and its distance from $y$, \\
$d_m^{2} - d_1^{2} + (z_1-R_{1,1}~x_1)^2,~ l_k \leftarrow l_k - \mathcal{I}_{\{x_k \neq 0\}}$ 
and go to 3. 
\\
\hline
\end{tabular}
\label{algo2}
\end{center}
\end{table}
\end{Huge}
\subsection{Computation of the lower bound $\mathcal{L}_k({\bf z}, \R, 
\x_{k:m}, \tilde l_k)$}
\label{comp_lower_bnd}

The technique described in this section leads to a reduction in the total number
of search tree nodes visited by the sparsity-aware sphere decoding algorithm.
However, this comes at an additional complexity incurred by computing 
the lower bound $\mathcal{L}_k({\bf z}, \R, \x_{k:m}, \tilde l_k)$ in (\ref{lowerbnd2}). 
Therefore, the benefit of additional pruning must outweigh the cost of
computing the aforementioned lower bound that enables the pruning.
Clearly, finding exact solution to the minimization problem in (\ref{lowerbnd2}) is 
equivalent to solving the original problem (\ref{firsteq}). As an alternative, we may 
relax the minimization in (\ref{lowerbnd2}) by introducing $\x \in \mathbb{R}^{k-1}$ 
and then solve
 \begin{eqnarray}
& \hspace{-0.2in}
 \underset{{\bf x} \in \mathbb{R}^{k-1} \atop \|\x\|_0 \le \tilde{l}_k}{\min} \|{\bf w}_{1:k-1} 
 - \R_{1:k-1,1:k-1}{\bf x}\|^2  
 &\label{lowerbnd7}
 \end{eqnarray}
Minimization (\ref{lowerbnd7}) can be solved using a number of techniques. To this
end, we rely on the computationally efficient orthogonal matching pursuit (OMP) 
method and round off each entry of the final solution to its nearest discrete value in 
$\mathcal{D}_L$. Details about the OMP algorithm are omitted for brevity and can be 
found in \cite{trop07}.

\subsection{Simulations}
\label{speedup_simulations}

\begin{figure*}
\begin{center}
\includegraphics[width=6.5in]{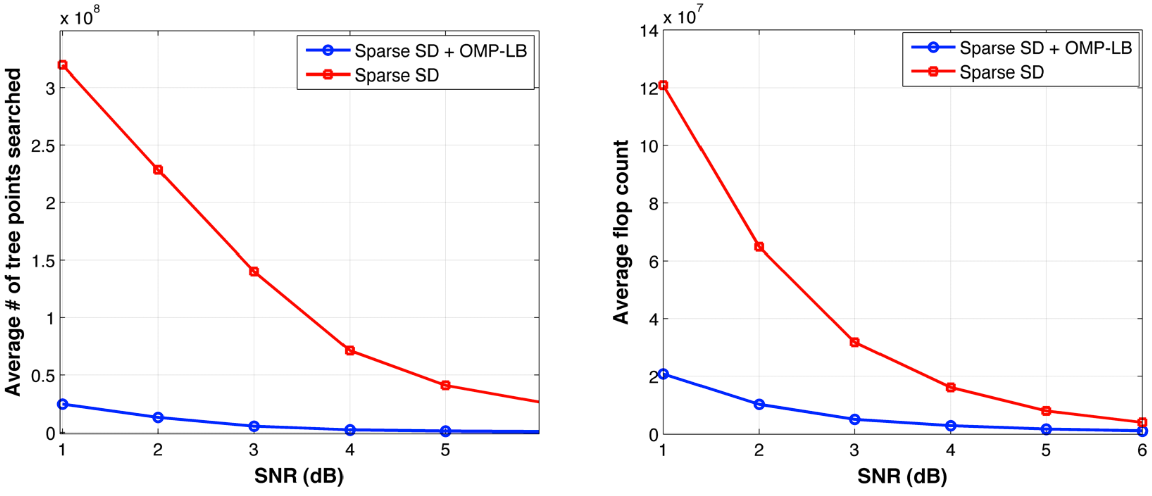}
\caption{\small Computational complexity of sparsity-aware sphere decoder with (blue) and without 
(red) lower bounding technique, $m = n = 40, l=5$, binary $\{0,1\}$ alphabet. The plot on the
left compares average number of the tree points visited by the algorithm and the plot on the right
compares average flop count.}
\label{sic_fast_sd}
\end{center}
\end{figure*}

We demonstrate the improvement in computational complexity of the modified 
sparsity-aware sphere decoding algorithm, where the lower bound in the relaxed problem (\ref{lowerbnd7}) 
is obtained by the OMP method. The results are shown in Figure \ref{sic_fast_sd} for 
$m=n=40$, $l = 5$, and binary $\{0,1\}$ alphabet. The subplot on the left compares
the average number of points in the tree searched by the two variants of the
algorithm (i.e., with and without using the lower-bounding technique). The subplot
on the right shows the total flop count on the average for the two methods. Clearly, for the considered 
range of parameters, the proposed modification of the algorithm enables significant 
computational savings over sparsity-aware sphere decoder from Section~II.

{\it Remark:} Note that the method presented in this section could benefit the complexity
of sparsity-aware sphere decoder at low to moderate SNRs, where the 
search sphere contains a large number of points and hence pruning tree nodes using the
considered lower bound may help. However, in high SNR regime, the sphere radius is so small 
that only few points lie inside the sphere and the cost of running OMP to find lower bounds 
may outweigh the benefit due to additional pruning.

\section{An Application: Sparse Channel Estimation via Sparsity-Aware Sphere Decoding}

In this section, we consider the application of the proposed sparsity-aware sphere decoding 
algorithm to the classical
problem of sparse channel estimation (see, e.g., \cite{mitra} and the references therein) and
benchmark its performance against state-of-the-art alternative techniques. In estimation of sparse channel, impulse response of a channel having a given delay spread is described by only few 
non-zero taps and the estimation goal is to determine the values of those taps. This problem is often 
encountered in digital television and underwater acoustic communications \cite{berger} where the
channel is naturally sparse, i.e., most of the channel energy is concentrated in few samples
of the impulse response. Sparse estimation also comes up in ultra-wideband communications where 
only dominant multipath components are estimated, leading to a sparse characterization of the 
channel. In \cite{mitra}, the sparse channel estimation is formulated as an On-Off Keying (OOK) 
problem by decoupling the estimation into zero-tap detection followed by structured estimation 
of the values of the non-zero taps. Sphere decoding based zero-tap detection is among several methodologies 
considered in \cite{mitra}. We show that our sparsity-aware sphere decoding algorithm enables
performance matching that of the best previously developed schemes, and comes close to the
fundamental performance limits.

\subsection{System model}

To learn the sparse channel impulse response, we transmit a known training sequence and
observe the received signal. Let $\{u_i\}_{i=1}^M$ denote the finite-length training data 
sequence and let $u_i = 0$ 
if $i > M$ or $i <0$. Assume that the channel has impulse response of length $L$ ( $>M$) and let $\h = [h_1~h_2~\ldots ~h_L ]^T$ denote the vector of the impulse 
response coefficients. 
The real-valued observations are collected into a vector $\x = [x_1~x_2~\ldots ~x_{M+L-1}]$,
and the relation between $\h$ and $\x$ is given by 
\begin{eqnarray}
& x_j = \sum\limits_{i=1}^{L}~h_i u_{j-i+1} + \nu_j, &  \nonumber 
\label{appl1}
\end{eqnarray}
for $j = 1, \ldots, M+L-1$, with $\nu_j$ assumed to be independent, identically distributed 
zero-mean Gaussian noise of variance $\sigma^2$. Alternatively, (\ref{appl1}) can be 
written as the matrix-vector product
\begin{equation}
\x = \U\h + \boldsymbol\nu,
\label{appl2}
\end{equation}
where 
\[
\U = 
\left[ \begin{array}{ccc}
u_1 &  & 0 \\
u_2 & \ddots &  \\
\vdots & \ddots & u_1 \\
u_M & & u_2 \\
 & \ddots & \vdots  \\
0 & & u_M 
 \end{array} \right] \mbox{ and }
\boldsymbol\nu = 
 \left[ \begin{array}{c}
\nu_1 \\
\nu_2 \\
\vdots \\
\nu_{M+L-1} \end{array} \right].
\]
Note that the channel vector can be represented as $\h = \mbox{diag}(h){\bf b}$, where 
${\bf b} \in \{0,1\}^L$ has non-zero entries that indicate positions of the non-zero 
coefficients in $\h$, and diag($h$) is a diagonal matrix with entries from $\h$. 

\subsection{Sparsity-aware sphere decoder for channel estimation}

Let $\hat{\h}$ denote the least-squares solution of (\ref{appl2}). Then the zero-tap detection problem 
can be formulated as the following optimization \cite{mitra} 
\begin{eqnarray}
&  \hat{{\bf b}} = \arg \underset{{\bf b} \in \{0,1\}^L}{\min} ~ \|\x - \U\mbox{diag}(\hat{h}){\bf b}\|^2. & 
\label{appl3}
\end{eqnarray} 
In \cite{mitra}, the sphere decoding algorithm was used to solve (\ref{appl3}) and find the point 
in the lattice $\U$diag($\hat{h}$) closest to $\x$. The solution $\hat{\bf b}$ is then used to
refine the channel estimate $\hat{\h}$ and form a so-called structured estimate $\hat{\h}^*$ 
via the least-squares method. As an alternative, we employ the sparsity-aware sphere decoding
from Section~\ref{sparseSD} to solve (\ref{appl3}). The channel order, denoted by $m^{\#}$, is
assumed known to the receiver. Then, (\ref{appl3}) can be restated as 
\begin{eqnarray}
&  \hat{{\bf b}} = \arg \underset{{\bf b} \in \{0,1\}^L \atop \|{\bf b}\|_0 ~\leq~ m^{\#}}{\min} ~ \|\x - 
\U\mbox{diag}(\hat{h}){\bf b}\|^2. & \label{appl4}
\end{eqnarray}

{\em Remarks}:
Various methods for estimating the channel order at the receiver exist in literature. In \cite{poor}, 
information theoretic criteria including Minimum Description Length (MDL) and Akaike Information 
Criterion (AIC) have been considered for determining the number of sources in sparse array 
processing \cite{array}. In \cite{perez}, a criterion for channel order estimation using least squares 
blind identification and equalization costs for SIMO channels was proposed. Often, information 
about the channel order is obtained from past history or data records. 

\subsection{Simulations} 
\begin{figure}[!t]
\centering
\includegraphics[width = 3.6in]{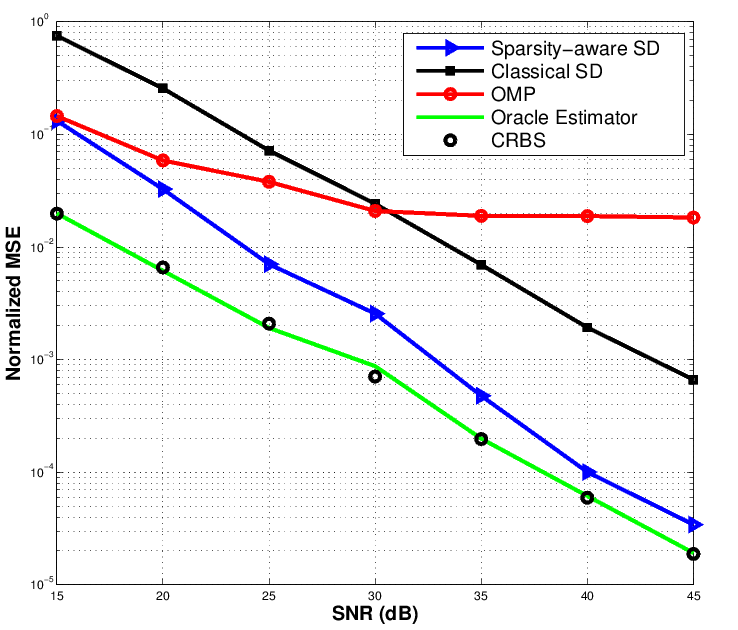}
\caption{\small Comparison of normalized mean-squared error (MSE) for sparsity-aware 
SD, OMP and classical SD schemes for zero-tap detection,  $L=20, M = 6,m^{\#} = 3$. The MSE of an oracle estimator and 
the CRLB for structured estimate are also plotted for comparison.}
\label{sparse_est}
\end{figure}

\begin{figure}[!t]
\centering
\includegraphics[width = 3.5in]{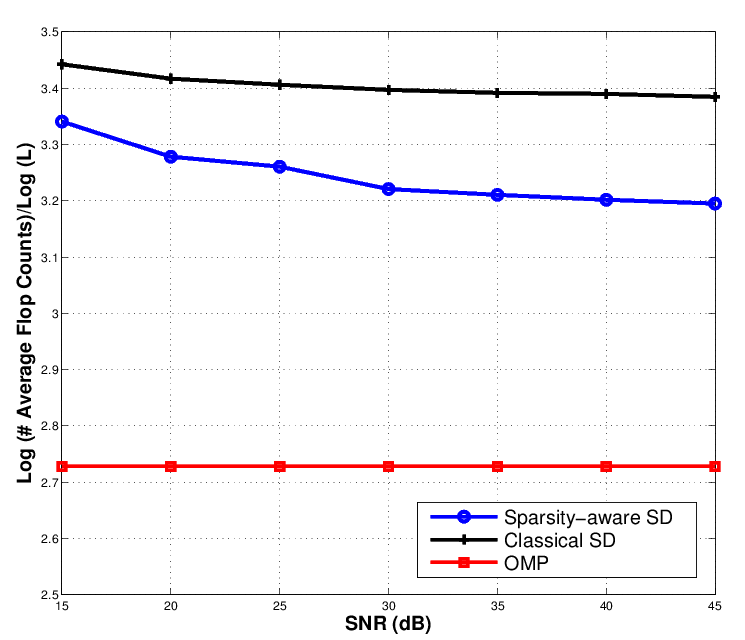}
\caption{\small Comparison of complexity exponents for average flop counts of sparsity-aware SD, OMP and classical SD schemes for zero-tap detection, $L = 20, M = 6, m^{\#} = 3$.}
\label{sparse_complexity}
\end{figure}

Performance of the sparsity-aware sphere decoding algorithm applied for sparse channel
estimation is demonstrated in Fig. \ref{sparse_est}, where we plot the normalized
mean-squared-error 
(MSE) of competing channel estimation schemes as a function of SNR. The normalized 
MSE is defined 
as MSE = $1/T\sum_{t =1}^{T} \|\h_t-\hat{\h}_t\|^2/\|\h_t\|^2$, where $T$ is the total 
number of Monte 
Carlo iterations and $\h_t$ and $\hat{\h}_t$ are the vectors of actual channel 
coefficients and 
their estimates in the $t^{th}$ iteration of the estimation procedure, respectively.  The values 
of $L$, $M$ and the channel order $m^{\#}$ are $20$, $6$ and $3$, respectively. The training 
sequence is formed by taking $M$ samples of constant modulus. The  non-zero coefficients 
of $\h_t$ are assumed to independent standard normal variables.

To solve the sparse channel estimation problem and obtain the MSE shown in 
Fig. \ref{sparse_est}, we employed the 
sparsity-aware sphere decoder from Section \ref{expected_complexity} with radius update
for faster run times. The MSE performance of the classical sphere decoding algorithm and structured estimate 
obtained via orthogonal matching pursuit (OMP) \cite{trop07} are among those shown in 
Fig \ref{sparse_est}. 
It is clear from this figure that sparsity-aware sphere decoder using 
knowledge of channel order performs significantly better than the classical sphere decoding 
adopted in \cite{mitra}. The MSE of the oracle estimator and the theoretical performance 
limit in the form of Cramer Rao lower bound (CRLB) \cite{mitra} are also plotted in the same 
figure. Oracle estimator is the structured least-square estimator with known locations of 
non-zero channel taps. In \cite{feng}, it was shown that OMP matches the best performance 
among sparse channel estimation methods that use knowledge of true channel order.
Clearly, over a wide range of signal-to-noise ratios (SNRs), the sparsity-aware 
sphere decoder significantly outperforms the OMP algorithm, with the performance gap 
increasing with SNR. 
In addition, we plot complexity exponents (evaluated
using the average flop counts incurred by the algorithms applied to sparse channel estimation
problem) in Fig. \ref{sparse_complexity}. It can be seen from this figure that sparsity-aware 
sphere decoding runs faster than the classical sphere decoder. The OMP algorithm is the 
fastest but its accuracy is inferior to sparsity-aware sphere decoding.


\section{Conclusion}

Sparsity-aware sphere decoder presented and analyzed in this paper has been motivated by 
the recent emergence of applications which require finding the solution to sparse 
integer least-squares problems. The sparsity-aware sphere decoding algorithm exploits 
sparsity information to prune nodes from the search tree that would have otherwise been 
visited by the classical sphere decoder. Imposing sparsity constraints requires minor 
additional computations, but due to more aggressive tree pruning, leads to significant 
improvements in complexity while improving accuracy. We analyzed computational complexity of 
the sparsity-aware sphere decoding algorithm and found its first and second-order 
moments (i.e., the mean and the variance of the random variable representing the complexity 
of the proposed algorithm). An application to sparse channel estimation demonstrated the 
efficacy of the algorithm.

To further reduce the computational complexity of the algorithm, we imposed a lower bound 
on a part of the objective function at each stage and used it to make the sphere constraint more
restrictive. The lower bound was posed as an $\ell_1$-norm regularized relaxed version of the
sparse integer least-squares problem, and efficiently computed using the orthogonal matching
pursuit technique. As simulation results demonstrate, the proposed method may lead to
significant computational savings over straightforward sparsity-aware sphere decoding.

As part of future work, in addition to first and second-order moments of the 
complexity of sparsity-aware sphere decoder analyzed in this manuscript, it would be 
beneficial to further characterize the complexity by, for example, exploring higher-order 
moments and its tail distribution. Moreover, for the lower-bound based speed-up of 
the algorithm, it is of interest to explore alternative relaxations of integer least-squares 
problems as well as investigate fast methods for solving such relaxations. Finally, applying 
the proposed techniques to solve various practical problems and exploiting structural 
properties of those problems to further improve the complexity and performance are of 
potential interest.


%

\section*{Acknowledgment}
The authors would like to thank Dr. Manohar Shamaiah for useful discussions. 
This work was in part supported by the NSF grant CCF \# 0845730.

\ifCLASSOPTIONcaptionsoff
  \newpage
\fi



\bibliographystyle{IEEEtran}
\bibliography{IEEEabrv,../bib/paper}
%
\appendix
\subsection{Proof of Lemma 1}
\label{lemma1proof}

We consider the two cases separately.
\\ \\
1) $k_1 < k_2$.

Define the support set of $\x_t^k$ as $S_j(\x_t^k) = \{i:\x_t^k(i) = j\}$, $j \in \{0,1\}$. 
Let us define 
\begin{eqnarray}
p ~=\!\!\! \sum\limits_{i \in S_1(\x_t^k)}^{}\!\!\!\! \! \! \mathcal{I}_{\{\x_a^k (i)=0\}}, & 
q ~=\!\!\! \sum\limits_{i \in S_0(\x_t^k)}^{}\!\!\!\! \! \! \mathcal{I}_{\{\x_a^k (i)=0\}}. &
\end{eqnarray}

In words, $p$ and $q$ denote the number of $0$'s in $\x_a^k$ in the indices where the entries of
$\x_t^k$ are equal to $1$ and $0$, respectively. Now, the $p$ zeros can be ordered over the 
support\footnote{Here, the support of a vector is the set of indices of its non-zero components.}
of $\x_t^k$ in $\binom{k_1}{p}$ different ways. For each such arrangement, the $q$ zeros can be 
ordered over the complement of the support of $\x_t^k$ (i.e., over $S_0(\x_t^k)$) in $\binom{k-k_1}{q}$ different ways. 
Therefore, the total number of ways of arranging $p$ and $q$ zeros in $\x_a^k$ over $1$'s and 
$0$'s of $\x_t^k$, respectively, is $\binom{k_1}{p}\binom{k-k_1}{q}$. The condition 
$\|\x_t^k-\x_a^k\|_0 =\eta$ implies that 
\vspace*{-1mm}
\begin{eqnarray}
\eta = p+(k-k_1-q) & \mbox{or} & p = (\eta-(k_2-k_1))/2, \nonumber
\end{eqnarray}
which is the desired result.
\\
2) $k_1 \geq k_2$.

Let us define
\begin{eqnarray}
p~= \!\!\! \sum\limits_{i \in S_1(\x_t^k)}^{}\!\!\!\! \! \! 
\mathcal{I}_{\{\x_a^k (i)=1\}}, & 
q~= \!\!\! \sum\limits_{i \in S_0(\x_t^k)}^{}\!\!\!\! \! \! 
\mathcal{I}_{\{\x_a^k (i)=1\}}. &
\end{eqnarray}
Here, $p$ and $q$ denote the number of $1$'s in $\x_a^k$ in the indices where 
the entries of $\x_t^k$ are equal to $1$ and $0$, respectively. The $q$ $1$'s can 
be ordered over the complement of the support of $\x_t^k$ in $\binom{k-k_1}{q}$ 
ways. For each such arrangement, the remaining $p$ $1$'s can be ordered over the 
support of $\x_t^k$ in $\binom{k_1}{p}$ ways. Therefore, the total number of ways 
of arranging $p$ and $q$ $1$'s in $\x_a^k$ over $1$'s and $0$'s of $\x_t^k$, 
respectively, is $\binom{k_1}{p}\binom{k-k_1}{q}$. The condition 
$\|\x_t^k-\x_a^k\|_0 =\eta$ implies that 
\begin{eqnarray}
\eta = (k_1-p)+q &\mbox{or}& q = (\eta-(k_1-k_2))/2, \nonumber
\end{eqnarray}
which is the desired result.

\subsection{Proof of Lemma 2}
\label{lemma2proof}
Recall that $p_{i,j}$ denotes the number of symbols $j$ 
in $\x_a^k$ in the positions where $\x_t^k$ has symbol $i$, and that $k_1 = \|\x_t^k\|_0$,
$k_2 = \|\x_a^k\|_0$. To enumerate all 
possible $l$-sparse vectors $\x_a^k$, we need to determine possible 
alignments of elements in $\x_t^k$ and $\x_a^k$. Define $p = p_{0,1} + p_{0,-1}$. 
We consider the following two cases separately.
\\ \\
1) $k_1 \leq k_2$.

The following observations can be made:

\begin{itemize}

\item[(a)] The number of ways in which $p_{1,0}$ $0$'s in $\x_a^k$ can be
aligned with components in $\x_t^k$ that are equal to $1$ is $\binom{a}{p_{1,0}}$, 
where $p_{1,0} \in \{0,1,\cdots,\min{(a,k-k_2)}\}$. Then, the number of ways
in which $p_{1,-1}$ $-1$'s in $\x_a^k$ can be aligned with the components in 
$\x_t^k$ that are equal to $1$ is $\binom{a-p_{1,0}}{p_{1,-1}}$, where 
$p_{1,-1} \in \{0, 1, \cdots, a-p_{1,0}\}$. Having fixed the positions of $0$'s and
$-1$'s, the remaining $p_{1,1} =(a-p_{1,0}-p_{1,-1})$ entries of $\x_a^k$
that are equal to $1$ can be aligned with the components in $\x_t^k$ that are 
equal to $1$ in only one way.

\item[(b)] Continuing the same type of arguments as above, the number of ways
in which $p_{-1,0}$ $0$'s in $\x_a^k$ can be aligned with components in $\x_t^k$ 
that are equal to $-1$ is $\binom{k_1-a}{p_{-1,0}}$, where 
$p_{-1,0} \in$ \\* $\{0, 1, \cdots, \min{(k_1-a,k-k_2- p_{1,0})}\}$. Following such an
arrangement, the number of ways in which $p_{-1,1}$ $1$'s in $\x_a^k$ can be 
aligned with the components in $\x_t^k$ that are equal to $-1$ is
$\binom{k_1-a-p_{-1,0}}{p_{-1,1}}$, where $p_{-1,1} \in \{0, 1, \cdots, k_1-a - p_{-1,0}\}$.
Once the positions of $0$'s and $1$'s are fixed, the remaining $p_{-1,-1} = 
k_1-a-p_{-1,1}-p_{-1,0}$ entries of $\x_a^k$ that are equal to $-1$ can be 
aligned with the components in $\x_t^k$ that are equal to $-1$ in only one way.

\item[(c)] 
Given (a), (b), the remaining $p = k_2-(k_1-p_{-1,0}-p_{1,0})$ non-zero entries of 
$\x_a^k$ can be aligned with $k-k_1$ zero-valued entries of $\x_t^k$ in 
$\binom{k-k_1}{k_2-(k_1-p_{-1,0}-p_{1,0})}$ ways. Since there are 2 types of non-zero 
symbol ($1$ and $-1$), there will be  $2^{k_2-(k_1-p_{-1,0}-p_{1,0})}$ combinations
in total.

\end{itemize}

Summarizing (a)-(c), we can enumerate vectors $\x_a^k$ (over parameters $k_1$, $k_2$
and $p_{i,j}$) as 
\begin{small}
\begin{align}
& g_1(k_1, k_2,k,p_{ij}) = \binom{a}{p_{1,0}}\binom{a-p_{1,0}}{p_{1,-1}} \binom{k_1-a}{p_{-1,0}}\binom{k_1-a-p_{-1,0}}{p_{-1,1}} 
\binom{k-k_1}{k_2-(k_1-p_{-1,0}-p_{1,0})}2^{k_2-(k_1-p_{-1,0}-p_{1,0})}, & \nonumber
\end{align} \end{small}
which is same as (\ref{ternary_case1}).
\\ \\
2) $k_1 > k_2$.

The following observations can be made:

\begin{itemize}

\item[(a)] The number of ways in which $p = p_{0,1} + p_{0,-1}$ non-zero elements
in $\x_a^k$ can be aligned with components in $\x_t^k$ that are equal to $0$ is 
$\binom{k-k_1}{p}$, where $p \in \{0, 1, \cdots, \min{(k_2, k-k_1)}\}$. Since there
are two non-zero symbols ($-1$ and $1$), there are $2^p$ ways to
arrange them. Having set positions and order of non-zero elements in $\x_a^k$,
the positions of $0$'s are uniquely determined. Next, we turn our attention to the
non-zero elements of $\x_t^k$.

\item[(b)] The number of ways in which $p_{1,-1}$ $-1$'s in $\x_a^k$ can be
aligned with components in $\x_t^k$ that are equal to $1$ is $\binom{a}{p_{1,-1}}$, 
where $p_{1,-1}=\{0, 1, \cdots, \min{(a,k_2- p)}\}$. Then, the number of ways
in which $p_{1,1}$ $1$'s in $\x_a^k$ can be aligned with the components in $\x_t^k$ 
that are equal to $1$ is $\binom{a-p_{1,-1}}{p_{1,1}}$, where 
$p_{1,1} \in \{(k_2-(k_1-a)-(p+p_{1,-1})), \cdots,  \min{(a - p_{1,-1}, k_2-(p+p_{1,-1}))}\}$.
Having fixed positions of $-1$'s and $1$'s, the remaining $p_{1,0}= (a-p_{1,-1} - p_{1,1})$ 
entries of $\x_a^k$ that are equal to $0$ can be aligned with the components in $\x_t^k$ 
that are equal to $1$ in only one way.

\item[(c)] Continuing the same type of arguments as above, the number of ways
in which $p_{-1,1}$ $1$'s in $\x_a^k$ can be aligned with components in $\x_t^k$ 
that are equal to $-1$ is $\binom{k_1-a}{p_{-1,1}}$, where  
$p_{-1,1} \in \{0, 1, \cdots,$ \\* $\min{(k_2-p-p_{1,-1}-p_{1,1}, k_1-a)}\}$. Following such 
an arrangement, the number of ways in which 
$p_{-1,-1} = k_2-(p+p_{1,-1}+p_{1,1} + p_{-1,1})$ $-1$'s in $\x_a^k$ can be aligned 
with the components in $\x_t^k$ that are equal to $-1$ is
$\binom{k_1-a-p_{-1,1}}{k_2-(p+p_{1,-1}+p_{1,1}+p_{-1,1})}$. Once the positions of 
$-1$'s and $1$'s are fixed, the remaining $p_{-1,0} = (k-k_2-p_{1,0}- p_{0,0})$
entries of $\x_a^k$ that are equal to $0$ can be aligned with the components in 
$\x_t^k$ that are equal to $-1$ in only one way.

\end{itemize}



Summarizing (a)-(c), we can enumerate vectors $\x_a^k$ (over parameters $k_1$, $k_2$
and $p_{i,j}$) as 
\begin{small}
\begin{align}
&g_2(k_1, k_2,k, p_{ij}) = \binom{k-k_1}{p_{0,1}+ p_{0,-1}}\binom{a}{p_{1,-1}} \binom{a-p_{1,-1}}{p_{1,1}}  \binom{k_1-a}{p_{-1,1}} 
\binom{k_1-a- p_{-1,1}}{k_2-(p_{0,1}+p_{0,-1}+p_{1,-1}+p_{1,1}+p_{-1,1})}2^{p_{0,1}+p_{0,-1}}, & \nonumber
\end{align} 
\end{small}
which is same as (\ref{ternary_case2}).

%

%






\end{document}